%% file: 1203.tex
\documentclass[a4paper,12pt]{article}
\usepackage{graphicx}
\usepackage[usenames]{color}
\newcommand{\la}{\langle}
\newcommand{\ra}{\rangle}

\newcommand{\beq}{\begin{eqnarray}}
\newcommand{\eeq}{\end{eqnarray}}

\newcommand{\bfg}{\mbox{{\boldmath $g$}}}

\newcommand{\bfs}{\mbox{{\boldmath $s$}}}
\newcommand{\bfG}{\mbox{{\boldmath $G$}}}
\newcommand{\bfu}{{\bf u}}

\newcommand{\bfU}{{\bf U}}

\newcommand{\bfF}{{\bf F}}
\newcommand{\bfW}{{\bf W}}

\newcommand{\bfC}{\mbox{{\boldmath $C$}}}
\newcommand{\bfR}{\mbox{{\boldmath $R$}}}

\newcommand{\bfrho}{\mbox{{\boldmath $\rho$}}}
\newcommand{\bfgamma}{\mbox{{\boldmath $\gamma$}}}

\newcommand{\eps}{\epsilon}

\newcommand{\e}{{\rm e}}

\renewcommand{\d}{\partial}

\renewcommand{\theequation}{\thesection.\arabic{equation}}

\newcommand{\PRL}{Phys.\ Rev. \ Lett.}

\def\Vec#1{\mbox{\boldmath$#1$}}

\begin{document}
\begin{center}
\begin{large}
{\bf Application of the Renormalization-group Method to the 
Reduction of Transport Equations}\\
\end{large}
\vspace{.5cm}
Teiji Kunihiro$^1$ and  Kyosuke Tsumura$^2$\\
\vspace{.3cm}
$^1$Yukawa Institute for
Theoretical Physics, Kyoto University,
Sakyo-ku, Kyoto 606-8502, Japan\\
$^2$Department of Physics, Kyoto University, Sakyo-ku, Kyoto 606-8502,
Japan
\end{center}


\begin{abstract}
We first give a comprehensive review of the renormalization group method
 for global and asymptotic analysis, putting an emphasis on the relevance
 to  the classical theory of envelopes and on the importance of
the existence of 
 invariant manifolds of the dynamics under consideration. 
We clarify that an essential point of the method is to  
 convert the problem from solving differential equations to obtaining
 suitable initial (or boundary) conditions:
The RG equation determines the slow motion
of the would-be integral constants in the unperturbative solution
 on the invariant manifold.
The RG method is applied to 
derive the Navier-Stokes equation from
the Boltzmann equation, as an example of the reduction of dynamics.
We work out to obtain the transport coefficients in terms of
the one-body distribution function.
\end{abstract}





\setcounter{equation}{0}
\renewcommand{\theequation}{\thesection.\arabic{equation}}
\section{Introduction}
The concept of the RG was introduced by 
Stuckelberg and Petermann as well as Gell-Mann and Low\cite{rg}
in relation to an ambiguity in the renomalization procedure of the
perturbation series in quantum field theory (QFT).
However, the essential nature  of the RG is exact and hence non-perturbative,
 which
was revealed and emphasized by Wilson\cite{wilson}.
Subsequently, as is well known, the machinery of the RG has been applied
to various problems in QFT and statistical physics with a great
success.\cite{rg}

The essence of the RG in quantum field theory (QFT) 
and statistical physics 
may be stated as follows: Let $\Gamma (\phi, {\bfg}(\Lambda), \Lambda)$
 be  the effective action (or thermodynamical potential) obtained
 by  integration of the field variable with the energy scale down to
 $\Lambda$ from infinity or a very large cutoff $\Lambda _0$.
Here ${\bfg}(\Lambda)$ is a collection of the coupling constants including
 the wave-function renormalization constant defined at
 the energy scale at $\Lambda$. Then
the RG equation may be expressed as a simple fact that
 the effective action as a functional of the field variable $\phi$
 should be the same,  irrespective to how much the integration of the
 field variable is achieved, i.e.,
\beq
\Gamma (\phi, {\bfg}(\Lambda), \Lambda)=
\Gamma (\phi, {\bfg}(\Lambda '), \Lambda ').
\eeq
If we take the limit $\Lambda '\rightarrow \Lambda$, we have
\beq
\label{eq:1-wilson}
\frac{d\Gamma(\phi, {\bfg}(\Lambda), \Lambda)}{d\Lambda}=0,
\eeq
which is the Wilson RG equation\cite{wilson}, or the 
flow equation in the Wegner's terminology \cite{weg}; notice
 that Eq.({\ref{eq:1-wilson})
is rewritten as 
\beq
\frac{\d\Gamma}{\d\bfg}\cdot\frac{d\bfg}{d\Lambda}=
-\frac{\d\Gamma}{\d\Lambda}. \label{eq:2-wilson}
\eeq
If the number of the coupling constants is finite, the theory is called
 renormalizable. In this case, the functional space of the theory does not
 change in the flow given by the variation of $\Lambda$.

Owing to the very non-perturbative nature,
the RG has at least two merits:
(A)~{\bf Resummation of the perturbation series}\,
Applying the RG equation of Gell-Mann-Low type \cite{rg} to 
perturbative calculations up to  first lowest orders,
a resummation in the infinite order of diagrams of some kind  can 
be achieved. That is, the RG method gives a powerful resummation 
method\cite{weinberg}.
(B)~{\bf Construction of infrared effective actions}\, 
The RG of Wilson type\cite{wilson} provides us
with a systematic method for constructing  low-energy 
effective actions which are asymptotically valid in the low-frequency
and long-wave length limit.

An appearance of diverging series
is a common phenomenon in every mathematical sciences not restricted
 in QFT, and some convenient ``resummation'' methods are
need and developed\cite{bo}.
Deducing a slow and long-wave length motion 
is one of the basic problems in almost all the fields of physics;
for example, statistical physics including 
the physics of the pattern formation and
the theories of collective motion in a many body 
system.
The problems may be collectively called the 
reduction problem of dynamics. 
The RG method\cite{cgo,kunihiro1,kunihiro2,kunihiro3,matu,efk,hatta}
 might be a unified method for the reduction of dynamics
as well as a powerful resummation method.

\begin{figure}[hp]
\begin{center}
  \includegraphics[width=0.3\textwidth,clip]{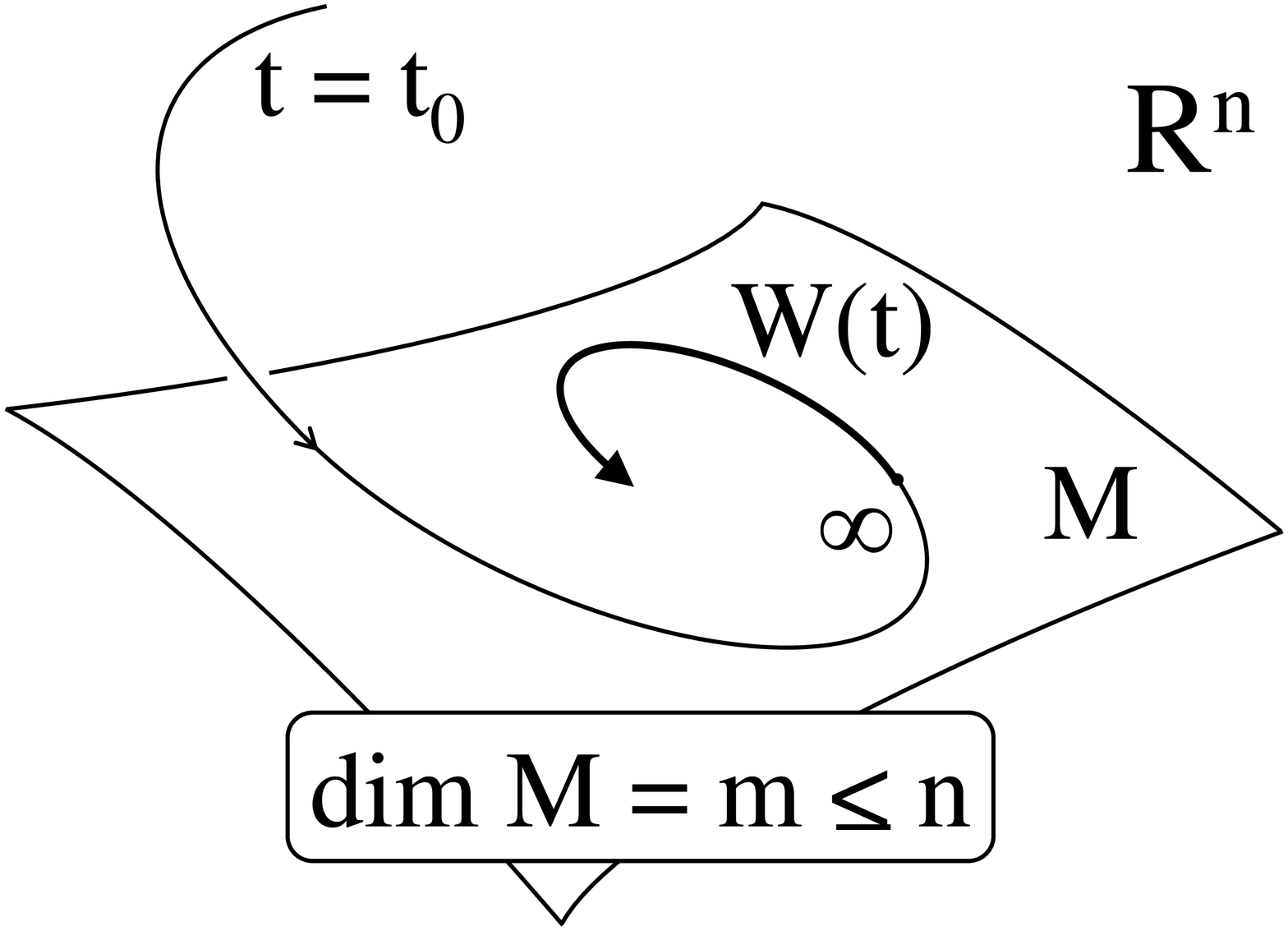}
\end{center}
  \caption[]{The geometrical image of the reduction of the dynamics.
The dynamical variable $\bfW(t)$ in the $n$-dimensional phase space 
approaches to and after some time  is eventually 
confined in the well-defined manifold M as  $t$
increases.  
}
  \label{fig:invmanifold}
\end{figure}

It is noteworthy that one can draw a clear geometrical image for  
the  reduction of  dynamical systems. Let $\bfW(t)$ be an
$n$-dimensional dynamical system governed by the evolution equation,
\beq
\frac{d \bfW}{dt}=\bfF(\bfW, t),
\eeq
where $\bfF$ is an $n$-dimensional vector; the dimension $n$ may be
finite or infinite. When the dynamics is reduced to an $m$-dimensional
system with $m$ being smaller than or equal to $n$, the vector $\bfW(t)$
approaches a well-defined $m$-dimensional manifold M embedded
in the $n$-dimensional phase space, as shown in Fig.1;
then the geometrical object M is called an attractive
manifold.  If after some time the 
$\bfW(t)$ is confined in the manifold M,
M is called an invariant  manifold. Furthermore,
 when the dynamics on M is slow, M is also called a slow manifold.
Let any point $\bfW$ within the manifold M be given by 
the relation $\bfW=\bfR(\bfs)$ with an $m$-dimensional parameter $\bfs(t)$.
Then the reduced dynamics is given by 
\beq
\label{reduceddynamics-exact}
\frac{d \bfs}{dt}=\bfG (\bfs), \quad \bfW =\bfR (\bfs),
\eeq
where the first equation with the vector field $\bfG(\bfs)$ defined on M 
gives the reduced dynamics within the manifold M
 and the second is the representation of M.
In the quantum field theory, the dynamical variable $\bfW$ corresponds
to the set of coupling constants and the renomalizability may be interpreted
as the existence of an invariant manifold in the space of coupling constants;
the increasing time $t$ and the vector field $\bfG(\bfs)$ correspond to 
the decreasing energy cutoff $\Lambda$ 
and the $\beta$ function, respectively.

In this article, 
(1)~ we show that the RG gives a powerful and systematic method for the reduction
of dynamics and also provides a transparent way for the construction
of the attractive slow manifold.
(2)~we apply the method to have the fluid dynamical limit of  the
Boltzmann equation as an example of the reduction of dynamics
and the construction of the slow manifold\cite{hatta}.
 We derive the Navier-Stokes equation explicitly
from the Boltzmann equation for the first time; the microscopic
expressions of the transport coefficients are given.
(3)~We will put an emphasis on the relation of the underlying
mathematics of the RG method with the classical theory of envelopes in
mathematical analysis
\cite{kunihiro1,kunihiro2,kunihiro3,matu,tanaka}.

\setcounter{equation}{0}
\section{The RG method and the classical theory of  envelopes}
 
We here give a brief review of the theory of  envelopes. 
Although the theory can be formulated in higher  
dimensions\cite{kunihiro1,kunihiro2,kunihiro3}, 
 we consider here envelope curves,
 for simplicity.

Let $\{C_{\tau}\}_{\tau}$ be a family of curves  parameterized by $\tau$ 
in the $x$-$y$  plane; here $C_{\tau}$ is  represented by the 
 equation $F(x, y, \tau)=0$.
 We suppose 
 that $\{C_\tau\}_{\tau}$ has the envelope $E$, which is represented by the
 equation $G(x, y)=0$, as shown in Fig.2.
The problem is to obtain $G(x, y)$ from $F(x, y,\tau)$.

Now let $E$ and a curve $C_{\tau_0}$ have the common tangent line at 
 $(x,y)=(x_0,y_0)$, i.e., $(x_0, y_0)$ is the point of tangency.
  Then $x_0$ and $y_0$ are functions of $\tau_0$;
$x_0=\phi(\tau_0),\ y_0= \psi(\tau_0)$, and of course $ G(x_0, y_0)=0$. 
Conversely, for each point $(x_0, y_0)$ on $E$, there exists a  parameter
 $\tau_0$.
 So we can reduce the problem to get $\tau_0$ as a function of 
$(x_0,y_0)$;
 then $G(x, y)$ is obtained as 
$F(x, y, \tau(x,y))=G(x,y)$.
Notice that since there
 is a relation $G(x_0, y_0)=0$ between $x_0$ and $y_0$, $\tau_0$ is actually a
 function of $x_0$ {\em or } $y_0$. $\tau_0(x_0, y_0)$ can be obtained as 
 follows.

Since the tangent line of $E$ at $(x_0, y_0)$  is perpendicular to the normal
 direction of $F(x, y, \tau)=0$ at the same point,  one has
$F_x(x_0, y_0, \tau_0)\phi'(\tau_0)+ F_y(x_0, y_0, \tau_0)\psi'(\tau_0)=0$.
 On the other hand, differentiating 
$F(x(\tau_0), y(\tau_0), \tau_0)=0$ with respect to $\tau_0$, one also has
$F_x(x_0, y_0, \tau_0)\phi'(\tau_0)+ F_y(x_0, y_0, \tau_0)\psi'(\tau_0)
 + F_{\tau_0}(x_0, y_0, \tau_0)=0$.
Combining the last two equations, we have
\beq
 F_{\tau_0}(x_0, y_0, \tau_0)
\equiv\frac{\partial F(x_0, y_0, \tau_0)}{\partial \tau_0}=0.
\eeq
This is the basic equation of the theory of envelopes.
Notice that the envelope equation has the similar 
form as the RG equation.
\begin{figure}[bph]
\begin{center}
  \includegraphics[width=0.35\textwidth,clip]{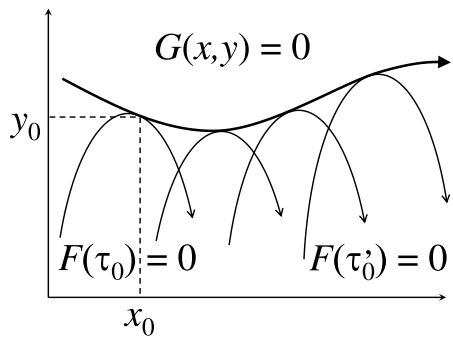}
\end{center}
  \caption[]{A family of curves $F(x, y, \tau_0)=0$ parameterized with
 $\tau_0$ and its envelope defined by $G(x,y)=0$.
}
  \label{envelope}
\end{figure}

One can thus 
 eliminate the parameter $\tau_0$ to get a relation between  $x_0$ and $y_0$;
$G(x, y) = F(x, y, \tau_0(x, y))=0$,
 with the replacement $(x_0, y_0)\rightarrow (x,y)$. $G(x, y)$ is called
 the discriminant of $F(x, y, t)$. 

When the function $F$ has an additional dependence on a
vector $\bfC (\tau)$, i.e., 
$F=F(x, y, \tau, \bfC(\tau))$, the envelope equation reads
\beq
F_{\tau_0}(x_0, y_0, \tau_0, \bfC(\tau_0))
\equiv\frac{\partial F(x_0, y_0, \tau_0)}{\partial \tau_0}
+\frac{\partial \bfC}{\partial \tau_0}
\frac{\partial F(x_0, y_0, \tau_0, \bfC(\tau_0))}{\partial \bfC}=0.
\eeq

Comments are in order here: 
(1)~When the family of curves is given by the
 function $ y=f(x, \tau)$, the envelope equation is reduced to 
${\partial f}/{\partial \tau_0}=0$; the envelope is given by 
 $y=f(x, \tau_0(x))$.\, 
(2)~The equation $G(x, y)=0$ may give 
 not only the envelope $E$ but also a set of 
singularities of the curves $\{C_{\tau}\}_{\tau}$.

\setcounter{equation}{0} 
\section{The RG method; a simplest example}

In this section, using a simplest example we show how the RG 
method  works for 
 obtaining  global and asymptotic behavior of solutions of differential
 equations. We shall present the method so that the reader 
will readily see that the 
notion of envelopes is 
intrinsically related to the method.
We shall emphasize that an essential point of the method is tuning the
 initial condition  at an arbitrary time $t_0$ perturbatively along with 
 solving the perturbative  equations successively. 
One will see that the reasoning for various steps in the procedure
 and the underlying picture are quite different from the
 original ones given in \cite{cgo}.
We believe, however, that the  present formulation  emphasizing the role of
initial conditions 
 and the relevance to envelopes of perturbative local solutions 
straightens the
 original argument, and is  the most 
comprehensive one.

Let us take the following simplest example to show 
our method:
\beq
\label{eq:do}
\frac{d^2 x}{dt^2}\ +\ \eps \frac{dx}{dt}\ +\ x\ =\ 0,
\eeq
where $\eps$ is supposed to be small. The solution to Eq.(\ref{eq:do}) reads
$ x(t)= \bar {A} \exp (-\frac{\eps}{2} t)\sin( \sqrt{1-\frac{\eps^2}{4}} t 
+ \bar {\theta})$,
where $\bar {A}$ and $\bar{\theta}$ are constants.

 Now, let us  obtain the solution around the initial time $t=t_0$ in a
 perturbative way,  expanding $x$ as
$
x(t, t_0) = x_0(t, t_0) \ +\ \eps  x_1(t ,t_0)\ +\ \eps ^2 x_2(t, t_0)\ 
+\ ... $,
 where $x_n(t, t_0)$ ($n= 0, 1, 2 ...$) satisfy
$\ddot{x}_0  +\ x_0\ =\ 0,
\ \ \ \
 \ddot{x}_{n+1}\ +\ x_{n+1}\ =- \dot{x}_n .$ 

The initial condition may be specified by
\beq
x(t_0, t_0)= W(t_0).
\eeq
We suppose that the initial value $W(t_0)$ is always on an exact solution of
 Eq.(\ref{eq:do}) for any  $t_0$. We also expand the initial value $W(t_0)$;
$W(t_0) = W_0(t_0) \ +\ \eps  W_1(t_0)\ +\ \eps ^2 W_2(t_0)\ 
+\ ... $,
and the terms $W_i(t_0)$ will be determined so that the perturbative solutions
 around different initial times $t_0$  have an 
 envelope.  Hence the initial value $W(t)$ thus constructed will give 
the (approximate but) global solution of the equation. 

Let us perform the above program. 
The lowest solution may be given by
$x_0(t, t_0) = A(t_0)\sin (t +\theta (t_0))$,
where we have made it explicit that the constants $A$ and $\theta$ may 
depend on the initial time $t_0$.
The initial value $W(t_0)$ as a function of $t_0$
 is specified as
$W_0(t_0)= x_0(t_0,t_0)= A(t_0)\sin (t_0 +\theta (t_0))$.
We remark that the zero-th order solution a neutrally stable solution; with 
 the perturbation $\eps \not=0$ 
 the constants $A$ and $\theta$ may move slowly.  We shall see that 
 the envelope equation gives the equations describing the slow motion of
 $A$ and $\theta$.
 
The first order equation now reads
$\ddot{x}_1\ +\ x_1\ =- A\cos(t+\theta),$
 and we choose the solution in the following form,
$x_1(t, t_0)= -\frac{A}{2}\cdot (t -t_0)\sin(t+\theta)$,
which means that the first order initial value $W_1(t_0)=0$ so that
the lowest order value $W_0(t_0)$ approximates the exact value as closely 
 as possible.
Similarly, the second order solution may be given by
$ x_2(t)= 
\frac{A}{8}\{ (t-t_0)^2\sin(t +\theta) - (t-t_0)\cos(t+\theta)\}$,
 thus $W_2(t_0)=0$ again for the present linear equation.

It should be noted  that  the secular terms have appeared 
 in the higher order terms, which are  absent in the 
exact solution and invalidates the perturbation theory for $t$ far
  from $t_0$. However, with the very existence of the secular terms,
 we could make $W_i(t_0)$ ($i=1, 2$) vanish and  $W(t_0)=W_0(t_0)$
 up to the third order.

Collecting the terms, we have 
\beq
\label{eq:all}
x(t, t_0)&=& A\sin (t +\theta) -\eps\frac{A}{2} (t -t_0)\sin(t+\theta)
  \nonumber \\ 
 \ \ \ & \ \ \ & +\eps^2\frac{A}{8}
\{ (t-t_0)^2\sin(t +\theta) - (t-t_0)\cos(t+\theta)\},
\eeq
and more importantly
$W(t_0)=W_0(t_0)=A(t_0)\sin (t_0 +\theta (t_0))$,
 up to $O(\eps^3)$. We remark that 
$W(t_0)$ describing the solution 
 is parameterized by possibly slowly moving variable $A(t_0)$ and 
 $\phi (t_0)\equiv t_0+\theta (t_0)$ in a definite way.

Now we have a family of curves $\{C_{t_0}\}_{t_0}$ given by functions 
$\{x(t, t_0)\}_{t_0}$ parameterized with $t_0$. 
 They are all on the exact curve $W(t)$ at $t=t_0$ 
 up to $O(\eps ^3)$, but 
 only valid locally for $t$ near $t_0$. 
 So it is conceivable that the envelope 
 $E$ of $\{C_{t_0}\}_{t_0}$ which 
 contacts with each local solution at $t=t_0$ will give a global solution.
 Thus the envelope function $x_{_E}(t)$ coincides with $W(t)$;  
$x_{_E}(t)=x(t,t)=W(t)$.

Our task is actually to determine $A(t_0)$ and $\theta(t_0)$ as 
 functions of $t_0$ so that the family of the local solutions has an 
 envelope.
 According to the classical theory of envelopes given in the previous
 section, the above program can be achieved by
 imposing that the envelope equation 
\beq
 \frac{dx(t, t_0)}{d t_0}=0,
\eeq
 gives the solution $t_0=t$.
From Eq.'s (\ref{eq:all}), we have
\beq
\frac{dA}{dt_0} + \eps A =0,  \ \ \ 
\frac{d\theta}{dt_0}+\frac{\eps^2}{8}=0,
\eeq
where we have used the fact that $dA/dt$ is $O(\eps)$ and neglected
 the terms of $O(\eps^3)$.
Solving the equations, we have 
$A(t_0)= \bar{A}{\rm e}^{-\eps t_0/2}$ and 
$\theta (t_0)= -\frac{\eps^2}{8}t_0 + \bar{\theta}$,
where $\bar{A}$ and $\bar{\theta}$ are constant numbers.
 Thus we get
\beq
x_{_E}(t)= x(t, t)=W_0(t)= 
\bar{A}\exp(-\frac{\eps}{2} t)\sin((1-\frac{\eps ^2}{8})t + 
\bar{\theta}),
\eeq
up to $O(\eps^3)$.
Noting that $\sqrt{1 - {\eps^2}/{4}}= 1 - {\eps^2}/{8} + O(\eps ^4)$, 
 one finds
 that the resultant envelope function $x_{_E}(t)=W_0(t)$ is an approximate but 
{\em  global} solution to Eq.(\ref{eq:do}).

\setcounter{equation}{0}
\section{The RG-reduction of dynamics of a generic evolution equation}
}

The theory of the reduction of evolution equations must give a definite
method to find out the vector field $\bfG(\bfs)$ as well as to construct the
attractive manifold M or the function $\bfR(\bfs)$ in 
Eq.(\ref{reduceddynamics-exact}).
It is often that these tasks can be achieved in a perturbative way
as follows,
\beq
\label{reduceddynamics}
\frac{d \bfs}{dt}=\bfG_0 (\bfs)+\bfgamma(\bfs), \quad 
\bfR(\bfs) =\bfR_0 (\bfs)+\bfrho(\bfs),
\eeq
where $\bfR_0 (\bfs)$ gives the coordinate of the unperturbed
invariant manifold
M$_0$ and $\bfrho$ gives the deformation of the manifold by the
perturbation, as shown in Fig.3.
The unperturbed vector field $\bfG_0(\bfs)$ governs
 the reduced dynamics on M$_0$ and 
$\bfgamma(\bfs)$ the modification of the dynamics. 
The important point lies in the fact that the modification of 
the manifold and the dynamics are both still function of
$\bfs$ parameterizing the unperturbed manifold M$_0$.

\begin{figure}[pbh]
\begin{center}
\includegraphics[width=0.3\textwidth,clip]{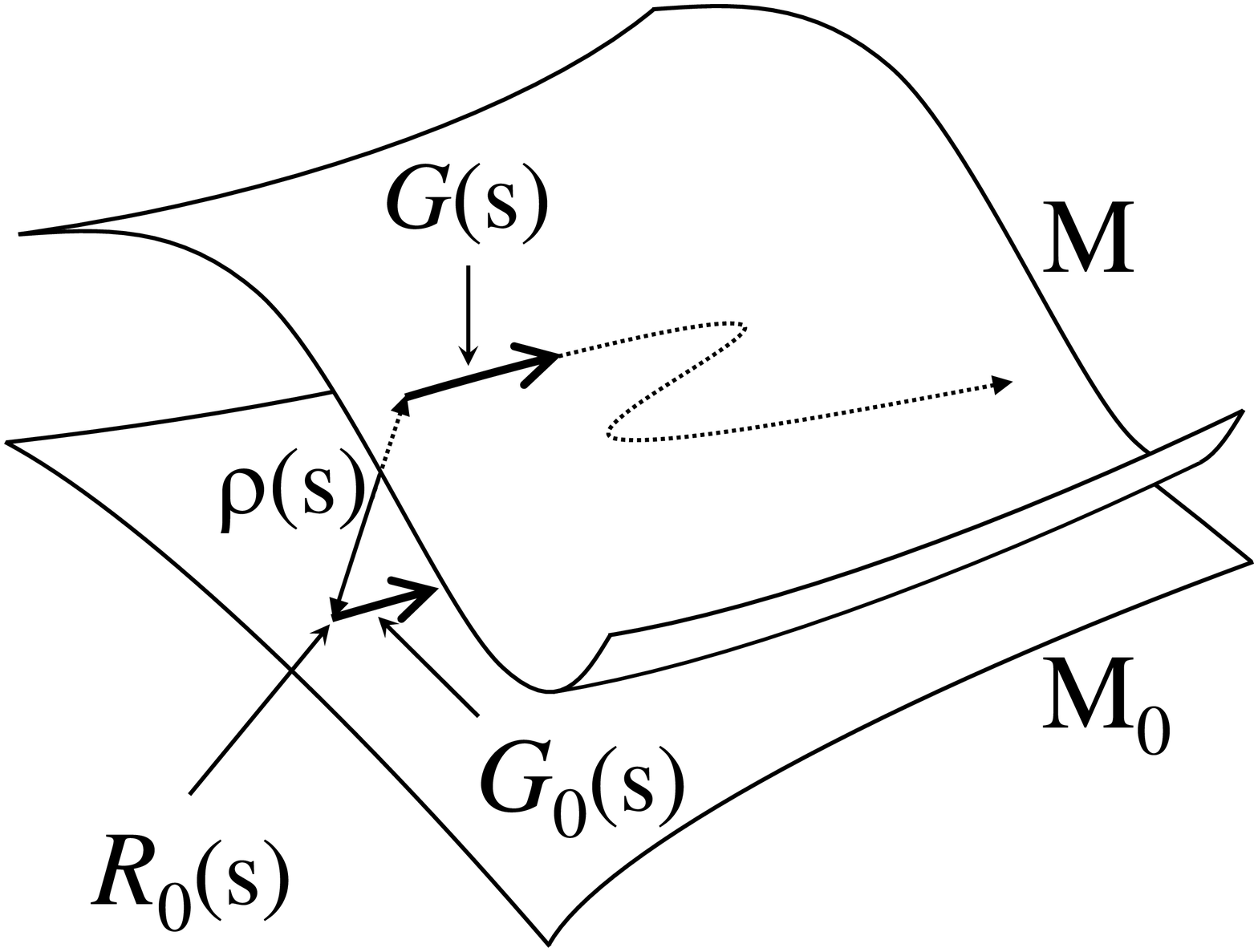}
\end{center}
  \caption[]{The geometrical image of the perturbative construction of 
the attractive manifold M and the reduced dynamics of the vector field
$\bfG$ on it. M$_0$ and $\bfG_0$ denote the unperturbed ones.
}
  \label{correspond}
\end{figure}

 Such a view on the reduction 
of dynamics was emphasized by Kuramoto\cite{kuramoto2}.
 It is remarkable that Bogoliubov
gave the notion of invariant manifold in his contribution to the theory
of non-linear oscillators\cite{krylov};
 he also described the fluid dynamical limit of 
Boltzmann equation as a construction of a (slow) invariant manifold spanned
by the hydrodynamical quantities inbeded in the functional space composed of
the sigle-particle distribution function\cite{bogo}.

In this section, we show 
how the renormalization-group method works to  make the reduction 
of the dynamics of a generic system possessing the possible reduction of
dynamics in the perturbative way\cite{efk}:
It will be clarified that the system reduction is accomplished
 by explicitly constructing the invariant manifold and 
the slow dynamics on the manifold in the perturbative way.
We emphasize that  the initial values are chosen
 by using a simple formula for the special solutions to differential equations
 as in the previous section.

We  treat the following rather generic  vector equations in this section:
\beq
\label{eq:ei-0}
\d _t\bfu =A\bfu+ \eps \bfF (\bfu),
\eeq
where $\d _t\bfu =\d \bfu/\d t$, 
$A$ is a linear operator, $\bfF$ a nonlinear function
of $\bfu$ and $\eps$ is a small parameter ($\vert \eps\vert<1$).
We assume that $A$ has multiply degenerated zero eigenvalues and 
other  eigenvalues of $A$ have a negative real part.
We assume that  $A$ has semi-simple $0$ eigenvalues in the present paper; 
the case when $A$ has a multidimensional Jordan cell is also
treated in \cite{efk}.

We are interested in constructing the attractive manifold M at 
$t\rightarrow \infty$ and the reduced dynamics on it.
We try to construct solve the problem
 in the perturbation theory by expanding $\bfu$ as
\beq
\bfu (t;t_0)= \bfu _0(t;t_0) +\eps \bfu_1(t;t_0) +\eps^2 \bfu_2(t;t_0) + 
\cdots,
\eeq
with the initial value $\bfW (t_0)$ at an arbitrary time $t_0$.
The equations in the first few orders read
\beq
\label{eq:ei-1}
(\d_t -A)\bfu_0=0,\quad
(\d_t-A)\bfu_1= \bfF(\bfu_0), \quad
(\d_t-A)\bfu_2&=& \bfF '(\bfu_0)\bfu_1, 
\eeq
where 
\beq
(\bfF '(\bfu_0)\bfu_1)_i=\sum_{j=1}^{n}
\left\{\d (F '(\bfu_0))_i/\d (u_0)_j\right\}(u_1)_j,
\eeq
is a Freche derivative for the $n$-dimensional vector.

We suppose that the equation has been solved up to $t=t_0$ and the solution
has the value $\bfW(t_0)$ at $t_0$.  Actually, the initial value must be 
determined by the perturbative solution self-consistently;
indeed, $\bfu(t)=\bfW(t)$ is the solution to (\ref{eq:ei-0}) in 
the global domain. Therefore it
should be also expanded as follows;
$\bfW(t_0)=\bfW_0(t_0)+\eps \bfW_1(t_0)+\eps^2\bfW_2(t_0)+\cdots = \bfW_0(t_0)+\bfrho(t_0)$,
where $\bfrho(t_0)$ is supposed to be an independent function of $\bfW_0$.
They are not yet known at present 
 but will be determined so that the perturbative
expansion becomes valid. One of the main purposes in this section 
is to show how sensibly the
initial values can be determined order by order.

%

In the present paper, we confine ourselves to the case where $A$ has  semi-simple 0
 eigenvalues. Let the dimension of ker$A$ be $m$;
$
A\bfU_i=0,$ ($i=1, 2, \dots ,m$).
We suppose that other eigenvalues have negative real parts;
$
A\bfU_{\alpha}=\lambda_{\alpha}\bfU_{\alpha},
 \quad (\alpha=m+1, m+2, \cdots , n)$,
where Re$\lambda_{\alpha}<0$.
One may assume without loss of generality 
that $\bfU_i$'s and $\bfU_{\alpha}$'s are linearly independent.

The adjoint operator $A^{\dag}$ has the same eigenvalues as $A$ has;
$
A^{\dag}\tilde{\bfU}_i=0, \quad (i=1, 2, \dots ,m)$
and
$A^{\dag}\tilde{\bfU}_{\alpha}= \lambda^{*}_{\alpha}\tilde{\bfU}_{\alpha}$,
$(\alpha =m+1, m+2, \cdots , n)$.
Here we suppose that $\tilde{\bfU}_i$' and $\tilde{\bfU}_{\alpha}$'s
 are linearly independent. Without loss of generality, one can choose
 the eigenvectors so that 
$
\la \tilde{\bfU}_i, \bfU_{\alpha}\ra=0=\la\tilde{\bfU}_{\alpha}, \bfU_i\ra$,
with $1\le i\le m$ and $m+1\le \alpha \le n$.
We denote  the projection operators by $P$ and $Q$ which projects onto
 the kernel of $A$ and the space orthogonal to Ker$A$, respectively.

Since we are interested in the asymptotic state as $t\rightarrow \infty$,
we  may assume that the lowest-order initial value  belongs to Ker$A$:
\beq
\bfW_0(t_0)=\sum_{i=1}^{m}C_i(t_0)\bfU_i=\bfW_0[\bfC].
\eeq
Thus trivially,
$\bfu_0(t;t_0)=\e^{(t-t_0)A}\bfW_0(t_0)=\sum_{i=1}^{m}C_i(t_0)\bfU_i$.
We notice that a natural parameterization of 
the invariant manifold in the lowest order M$_0$ is 
given by the set of the integral constants 
$\bfC=\ ^t(C_1, C_2, \cdots, C_m)$ being varied.

The first order equation (\ref{eq:ei-1}) 
with the initial value $\bfW_1(t_0)$ 
 which is not yet determined is formally 
solved to be
\beq
\bfu_1(t;t_0)&=&
\e^{(t-t_0)A}[\bfW_1(t_0)+A^{-1}Q\bfF(\bfW_0(t_0))]\nonumber \\
            & & +(t-t_0)P\bfF (\bfW_0(t_0))-A^{-1}Q\bfF(\bfW_0(t_0)).
\eeq
The first term has a possibility to give rise to a fast motion, which should 
be avoided  and  are
analogous to 
 divergent terms in quantum field theory; the divergent terms
 are subtracted away by 
  counter terms, which are analogue to the initial values $\bfW_i$ here.
Indeed it is nice that the initial value $\bfW_1(t_0)$ not yet determined 
can be chosen so as to cancel out the would-be fast term
 as follows;
$\bfW_1(t_0)=-A^{-1}Q\bfF(\bfW_0(t_0))$,
which satisfies $P\bfW_1(t_0)=0$ and is a function solely of $\bfC(t_0)$.
  Thus we have for the first order solution,
$\bfu_1(t;t_0)=(t-t_0)P\bfF -A^{-1}Q\bfF$,
where the argument of $\bfF$ is $\bfW_0[\bfC]$.
Now the invariant manifold is modified to M$_1$ given by
\beq
{\rm M}_1=\{\bfu \vert \bfu=\bfW_0-\eps A^{-1}Q\bfF(\bfW_0)\}.
\eeq

If one stops to this order, the approximate solution reads
\beq
\bfu(t;t_0)=\bfW_0+\eps\{
    (t-t_0)P\bfF -A^{-1}Q\bfF
                            \}.
\eeq
Then the RG equation $\d\bfu/\d t_0\vert_{t_0=t}=0$ gives 
$\dot{\bfW}_0(t)=\eps P\bfF(\bfW_0(t))$,
which is reduced to an $m$-dimensional coupled equation,
\beq
\label{eq:rg-gen-2}
\dot{C}_i(t)=\eps \la\tilde{\bfU}_i, \bfF(\bfW_0[\bfC])\ra,
 \quad (i=1, 2, \cdots ,m).
\eeq
One now sees that $\eps P\bfF(\bfW_0[\bfC])$ gives the vector filed
$\bfG_0(\bfs)$ with $\bfC$ being identified with $\bfs$ in
 Eq.(\ref{reduceddynamics}). 
The global solution representing a trajectory on the invariant manifold 
up to this order  is given by                              
\beq
\label{eq:rg-sol-1}
\bfu(t)=\bfu(t; t_0=t)=\sum_{i=1}^{m}C_i(t)\bfU_i
- \eps A^{-1}Q\bfF(\bfW_0[\bfC]),
\eeq
with $\bfC(t)$ being the solution to (\ref{eq:rg-gen-2}).

In short, we have derived the invariant manifold as 
the initial value represented by (\ref{eq:rg-sol-1})
 and the reduced dynamics (\ref{eq:rg-gen-2}) on it in the RG
  method in the first order 
 approximation.
 
The above procedure can be easily extended to second and higher orders
and the modification of the vector field $\bfgamma(\bfs)$ in
Eq.(\ref{reduceddynamics}) is readily obtained,
as shown in \cite{efk}.

\setcounter{equation}{0}
\section{Fluid dynamical limit of Boltzmann equation}  
 
In this section, we apply the RG method formulated in the previous
sections and in \cite{efk} to obtain the fluid dynamical limit of the Boltzmann
equation\cite{re:boltzmann}.
 This is an example of reducing a kinetic
equation to a slower dynamics\cite{hatta}.
In this article, we complete the derivation of the Navier-Stokes equation
with some corrections to the previous treatment\cite{hatta}; we shall also 
work out to give the explicit formulae of the transport coefficients.

 \subsection{Basics of the Boltzmann Equation}
 
 The Boltzmann equation\cite{re:boltzmann, resibois} is a
 transport equation which describes the time evolution of one-particle
 distribution function defined in the phase space:
 \begin{eqnarray}
  \frac{\partial}{\partial t}f(\Vec{r} ,\, \Vec{v} ,\, t) +
   \Vec{v}\cdot\Vec{\nabla}f(\Vec{r} ,\, \Vec{v} ,\, t) = I[f](\Vec{r} ,\,
   \Vec{v} ,\, t)\label{eq:001}.
 \end{eqnarray}
 The right-hand side of the above equation is called the collision integral, 
 {\small
 \begin{eqnarray}
   I[f](\Vec{r} ,\, \Vec{v} ,\, t) =
   \int\!\!\Vec{d^3v_1}\int\!\!\Vec{d^3v_2}\int\!\!\Vec{d^3v_3} \,
   \omega(\Vec{v} ,\, \Vec{v_1}|\Vec{v_2} ,\, \Vec{v_3}) \, \Big(
   f(\Vec{r} ,\, \Vec{v_2} ,\, t)f(\Vec{r} ,\, \Vec{v_3} ,\, t) -
   f(\Vec{r} ,\, \Vec{v} ,\, t)f(\Vec{r} ,\, \Vec{v_1} ,\, t) \Big)\label{eq:002},
 \end{eqnarray}
 }
 where $\omega(\Vec{v} ,\, \Vec{v_1}|\Vec{v_2} ,\, \Vec{v_3})$ denotes
 the transition probability which comes from a microscopic two-particle
 interaction.
 We remark that the transition probability $\omega(\Vec{v} ,\,
 \Vec{v_1}|\Vec{v_2} ,\, \Vec{v_3})$ contains delta functions reflecting
 energy-momentum conservation law, 
 and satisfies the following relations based on the indistinguishability
 of identical particles and the time reversal symmetry in the scattering
 process:
 $
 \omega(\Vec{v} ,\, \Vec{v_1}|\Vec{v_2} ,\, \Vec{v_3}) =
 \omega(\Vec{v_1} ,\, \Vec{v}|\Vec{v_3} ,\, \Vec{v_2}) = 
 \omega(\Vec{v_2} ,\, \Vec{v_3}|\Vec{v} ,\, \Vec{v_1}) = 
 \omega(\Vec{v_3} ,\, \Vec{v_2}|\Vec{v_1} ,\, \Vec{v}).
 $
 
 To make explicit the correspondence to the general formulation given in
the previous section\cite{efk}, one may treat the argument $\Vec{v}$ as a
 discrete variable\cite{kuramoto2}. Discriminating the arguments $(\Vec{r}
 ,\, t)$ and $\Vec{v}$, we use $\Vec{v}$ as a subscript for the
 distribution function:
$  f(\Vec{r} ,\, \Vec{v} ,\, t) = f_{\Vec{v}}(\Vec{r} ,\, t) \equiv
   \big[ \Vec{f}(\Vec{r} ,\, t) \big]_{\Vec{v}}$.
 Then the Boltzmann equation now reads
 \begin{eqnarray}
  \frac{\partial}{\partial t}f_{\Vec{v}}(\Vec{r} ,\, t) +
   \Vec{v}\cdot\Vec{\nabla}f_{\Vec{v}}(\Vec{r} ,\, t) =
   I[f]_{\Vec{v}}(\Vec{r} ,\, t)\label{eq:006},
 \end{eqnarray}
 where
\begin{eqnarray}
  I[f]_{\Vec{v}}(\Vec{r} ,\, t) \equiv
   \sum_{\Vec{v_1}}\sum_{\Vec{v_2}}\sum_{\Vec{v_3}} \, \omega(\Vec{v} ,\,
   \Vec{v_1}|\Vec{v_2} ,\, \Vec{v_3}) \, \big( f_{\Vec{v_2}}(\Vec{r} ,\,
   t)f_{\Vec{v_3}}(\Vec{r} ,\, t) - f_{\Vec{v}}(\Vec{r} ,\,
   t)f_{\Vec{v_1}}(\Vec{r} ,\, t) \big).
\label{eq:007}
 \end{eqnarray}
 
 As promised, we apply the RG method
 to extract the low-frequency dynamics from a given kinetic equation;
 in other words, we achieve the coarse-graining of temporal scale by the RG
 method.
 We introduce $\epsilon$ in front of spatial derivative of the Boltzmann
 equation to make the application of the perturbation theory possible:
 \begin{eqnarray}
  \frac{\partial}{\partial t}f_{\Vec{v}}(\Vec{r} ,\, t) =
   I[f]_{\Vec{v}}(\Vec{r} ,\, t) - \epsilon \,
   \Vec{v}\cdot\Vec{\nabla}f_{\Vec{v}}(\Vec{r} ,\, t)\label{eq:008}.
 \end{eqnarray}
 
 \subsection{Procedure 1 : invariant manifold \& approximate solution}
  
  As was done in the previous sections,
 we first expand the initial values as  follows:
 \begin{eqnarray}
   f_{\Vec{v}}(\Vec{r} ,\, t_0) = 
    f^{(0)}_{\Vec{v}}(\Vec{r} ,\, t_0) + \epsilon \, 
    f^{(1)}_{\Vec{v}}(\Vec{r} ,\, t_0) + \epsilon^2 \,
    f^{(2)}_{\Vec{v}}(\Vec{r} ,\, t_0) + \cdots.
\label{eq:009} 
\end{eqnarray}
  Then let $\tilde{f}_{\Vec{v}}(\Vec{r} ,\, t \,;\, t_0)$ be an
  approximate solution around $t=t_0$,
 which 
 obeys Eq.(\ref{eq:008}) with the initial
condition at $t = t_0$:
$\tilde{f}_{\Vec{v}}(\Vec{r} ,\, t_0 \,;\, t_0) = f_{\Vec{v}}(\Vec{r} ,\, t_0)$.
  We try to solve $\tilde{f}_{\Vec{v}}(\Vec{r} ,\, t \,;\, t_0)$
by the perturbation theory by expanding it as
\begin{eqnarray}
   \tilde{f}_{\Vec{v}}(\Vec{r} ,\, t \,;\, t_0) =
    \tilde{f}^{(0)}_{\Vec{v}}(\Vec{r} ,\, t \,;\, t_0) + \epsilon \,
    \tilde{f}^{(1)}_{\Vec{v}}(\Vec{r} ,\, t \,;\, t_0) + \epsilon^2 \,
    \tilde{f}^{(2)}_{\Vec{v}}(\Vec{r} ,\, t \,;\, t_0) + \cdots,
\label{eq:011}
  \end{eqnarray}
with the respective initial conditions;
  $
   \tilde{f}^{(\mu)}_{\Vec{v}}(\Vec{r} ,\, t_0 \,;\, t_0) =
    f^{(\mu)}_{\Vec{v}}(\Vec{r} ,\, t_0) \,\,\, \mathrm{for} \,\,\, \mu = 0 ,\,
    1 ,\, 2 ,\, \cdots.
    $
  Substituting the above expansion in (\ref{eq:008}), we obtain the
  series of the perturbative equations. The lowest few equations read
  \begin{eqnarray}
   \frac{\partial}{\partial t}\tilde{f}^{(0)}_{\Vec{v}}
 &=& I[f]_{\Vec{v}}\Bigg|_{\Vec{f} =
    \tilde{\Vec{f}}^{(0)}}\label{eq:013},\\
   \frac{\partial}{\partial t}\tilde{f}^{(1)}_{\Vec{v}} &=& \sum_{\Vec{k}} \,
    \frac{\partial}{\partial f_{\Vec{k}}}I[f]_{\Vec{v}}\Bigg|_{\Vec{f} =
    \tilde{\Vec{f}}^{(0)}} \, \tilde{f}^{(1)}_{\Vec{k}} -
    \Vec{v}\cdot\Vec{\nabla}\tilde{f}^{(0)}_{\Vec{v}}\label{eq:014},\\
   \frac{\partial}{\partial t}\tilde{f}^{(2)}_{\Vec{v}} &=& \sum_{\Vec{k}} \,
    \frac{\partial}{\partial f_{\Vec{k}}}I[f]_{\Vec{v}}\Bigg|_{\Vec{f} =
    \tilde{\Vec{f}}^{(0)}} \, \tilde{f}^{(2)}_{\Vec{k}} - 
    \Vec{v}\cdot\Vec{\nabla}\tilde{f}^{(1)}_{\Vec{v}}\label{eq:015}.
  \end{eqnarray}
  
  Here  a remark is in order: We have actually used the
  linearized Boltzmann equation \cite{resibois} 
neglecting the second-order term of
  $\tilde{\Vec{f}}^{(1)}$ in (\ref{eq:015}). It is known that the
  neglected term produces the so-called Burnett terms which are absent in
  the usual Navier-Stokes equations\cite{re:burnett}.
  
  \subsection{Procedure 2 : order-by-order analysis}
  
Here we summarize the significance  of each order 
of equation:
  \begin{itemize}
   \item{0th order}\\
	The 0th-order equation is in general a nonlinear algebraic equation
	which determines the lowest-order invariant manifold $M_0(\equiv
	\tilde{\Vec{f}}^{(0)})$. The special solution of this equation
	includes some integration constants. It turns out that
      these would-be constants become 
    the slow variables on $M_0$ by the RG equation.
   \item{1st order}\\
	The 1st-order equation is a linear differential equation because
	we consider the dynamics near $M_0$. First we derive the zero modes
	from the eigen vectors of its time evolution operator.
	Then we define the appropriate inner product and
	projection operator to the kernel space spanned by the zero modes.
	Using these definitions, we solve the linear differential
	equation. The condition that the fast modes orthogonal to the zero
	modes vanish determine the initial condition in this order
         and thereby the 1st-order invariant manifold
	$M_1(\equiv\tilde{\Vec{f}}^{(1)})$; the 1st-order perturbation
	gives the deformation of the invariant manifold from $M_0$ to
   $M_1$. The coordinates to describe the slow modes are still defined
       on $M_0$.
   \item{2nd and higher orders}\\
	The 2nd-order equation is a linear differential equation with
	the same time evolution operator. We can solve the 2nd-order
	equation with the same procedure as the 1st one. The 2nd-order
	invariant manifold $M_2(\equiv\tilde{\Vec{f}}^{(2)})$ is
	determined in the same manner as in the 1st order. This procedure is
	able to be continued up to  arbitrary orders. Although 
	the successive deformation of the invariant manifold is obtained,
	the coordinates for the slow variables are still on
	$M_0$.
  \end{itemize}
  
   Now we are interested in the slow motion which may be realized
   asymptotically as $t \rightarrow \infty$. Therefore we put the
   stationary condition on (\ref{eq:013}) and the 0th-order equation
   becomes the following nonlinear algebraic equation:
   \begin{eqnarray}
    \frac{\partial}{\partial t}\tilde{f}^{(0)}_{\Vec{v}} = 0
     \,\,\,\Longrightarrow\,\,\, I[f]_{\Vec{v}}\Bigg|_{\Vec{f} = \tilde{\Vec{f}}^{(0)}}
     = 0\label{eq:016}.
   \end{eqnarray}
   The 0th-order approximate solution $\tilde{\Vec{f}}^{(0)}$ is the fixed
   point of the collision integral, and thus is a local equilibrium
   distribution function or Maxwellian:
   \begin{eqnarray}
    \tilde{f}^{(0)}_{\Vec{v}}(\Vec{r} ,\, t \,;\, t_0) = n(\Vec{r} ,\, t_0)
     \, \Bigg[ \frac{m}{2\pi T(\Vec{r} ,\, t_0)} \Bigg]^{\frac{3}{2}} \,
     \exp\Bigg[ - \frac{m|\Vec{v} - \Vec{u}(\Vec{r} ,\, t_0)|^2}{2T(\Vec{r}
     ,\, t_0)} \Bigg] \equiv f^{\mathrm{eq}}_{\Vec{v}}(\Vec{r} \,;\, t_0)\label{eq:017}.
   \end{eqnarray}
   Here, the local density $n(\Vec{r} ,\, t_0)$, 
local temperature $T(\Vec{r} ,\, t_0)$, and local flux
   $\Vec{u}(\Vec{r} ,\, t_0)$ are all dependent 
on the initial time $t_0$ and the space
   coordinate $\Vec{r}$ but independent of time $t$.
   The 0th-order invariant manifold $\Vec{f}^{(0)}$ is given by the
   initial condition:
   $
    f^{(0)}_{\Vec{v}}(\Vec{r} ,\, t_0) = \tilde{f}^{(0)}_{\Vec{v}}(\Vec{r}
     ,\, t_0 \,;\, t_0) = f^{\mathrm{eq}}_{\Vec{v}}(\Vec{r} \,;\, t_0).
     $
   The 0th-order result is summarized as follows:
   \begin{eqnarray}
    \left\{
     \begin{array}{l}
      \displaystyle{\tilde{\Vec{f}}^{(0)}(t) = \Vec{f}^{\mathrm{eq}}},\\[0.2cm]
      \displaystyle{\Vec{f}^{(0)}(t_0) = \Vec{f}^{\mathrm{eq}}}.
     \end{array}
    \right.\label{eq:019}.
   \end{eqnarray}
   
   The 1st-order equation now  reads
   \begin{eqnarray}
    \frac{\partial}{\partial t}\tilde{\Vec{f}}^{(1)} = A \,
     \tilde{\Vec{f}}^{(1)} + \Vec{F}\label{eq:020},
   \end{eqnarray}
   where the time evolution operator $A$ and the inhomogeneous term
   $\Vec{F}$ are defined by
   \begin{eqnarray}
    A_{\Vec{v}\Vec{k}} \equiv \frac{\partial}{\partial
     f_{\Vec{k}}}I[f]_{\Vec{v}}\Bigg|_{\Vec{f} = \Vec{f}^{\mathrm{eq}}},\,\,\,\,\,\,
     F_{\Vec{v}} \equiv - \Vec{v}\cdot\Vec{\nabla}f^{\mathrm{eq}}_{\Vec{v}}\label{022},
   \end{eqnarray}
respectively.
   
   As mentioned above, we must clarify the properties of the
   linear operator $A$ to proceed further. For this
   purpose, we convert $A$ to the following operator,
   \begin{eqnarray}
    L_{\Vec{v}\Vec{k}} \equiv f^{\mathrm{eq}\mbox{-}1}_{\Vec{v}} \,
     A_{\Vec{v}\Vec{k}} \, f^{\mathrm{eq}}_{\Vec{k}} =
     - \sum_{\Vec{v_1}}\sum_{\Vec{v_2}}\sum_{\Vec{v_3}} \, \omega(\Vec{v} ,\,
     \Vec{v_1}|\Vec{v_2} ,\, \Vec{v_3}) \, f^{\mathrm{eq}}_{\Vec{v_1}} \,
     \Big( \delta_{\Vec{v}\Vec{k}} + \delta_{\Vec{v_1}\Vec{k}} -
     \delta_{\Vec{v_2}\Vec{k}} - \delta_{\Vec{v_3}\Vec{k}} \Big)\label{eq:023},
   \end{eqnarray}
   which is called the collision operator\cite{chapman, kawasaki}.
   
Let us define the inner product between arbitrary
   vectors, $\Vec{\varphi}$ and $\Vec{\psi}$ by
   \begin{eqnarray}
  \langle \, \Vec{\varphi} \,,\, \Vec{\psi} \, \rangle \equiv \sum_{\Vec{v}} \,
 f^{\mathrm{eq}}_{\Vec{v}} \, \varphi_{\Vec{v}} \, \psi_{\Vec{v}}\label{eq:024}.
   \end{eqnarray}
We remark that this definition of the inner product is more adequate
than that given in \cite{hatta}.
A nice point is that $L$ becomes self-adjoint with this inner product;
   $\langle \, \Vec{\varphi} \,,\, L \, \Vec{\psi} \, \rangle = \langle
   \, L \, \Vec{\varphi} \,,\, \Vec{\psi} \, \rangle$.
 It is essential for the following discussions that
$L$ has the zero modes and dim[Ker$L$]$=5$\cite{resibois}:
\beq
  L \, \Vec{\varphi}^{0}_{\alpha} = \Vec{0},\quad 
(\alpha=0, 1,2,3,4)
\eeq
where the normalized five vectors $\Vec{\varphi}^{0}_{\alpha}$
are given as follows with
$ \Vec{\delta v} \equiv \Vec{v} - \Vec{u}$;
   \begin{eqnarray}
    \varphi^{0}_{0 \, \Vec{v}} &\equiv& \frac{1}{\sqrt{n}}\label{eq:027},\\
    \varphi^{0}_{i \, \Vec{v}} &\equiv& \frac{1}{\sqrt{n}} \,
     \sqrt{\frac{m}{T}} \, \delta v^i \,\,\, \mathrm{for} \,\,\, i = 1
     ,\, 2 ,\, 3\label{eq:028},\\
    \varphi^{0}_{4 \, \Vec{v}} &\equiv& \frac{1}{\sqrt{n}} \,
     \sqrt{\frac{2}{3}} \, \Big( \frac{m}{2T} \, |\Vec{\delta v}|^2 -
     \frac{3}{2} \Big)\label{eq:031},.
   \end{eqnarray}
with
$\langle \, \Vec{\varphi}^0_{\alpha} \,,\, \Vec{\varphi}^0_{\beta} \, \rangle =
     \delta_{\alpha\beta}$.
   The other eigenvalues are found to be negative; in fact one can show
that $\langle \, \Vec{\varphi} \,,\, L \, \Vec{\varphi} \, \rangle \le 0
   \,\,\, \mathrm{for} \,\,\, \mathrm{all} \,\,\, \Vec{\varphi}$,
which means that the kinetic dynamics near the 0th-order solution
 $\Vec{f}^{\mathrm{eq}}$ has an attractive slow manifold.
   
   Next we define the following projection operator $P$ onto the kernel
 of $L$,
   \begin{eqnarray}
    \Big[ P \, \psi \Big]_{\Vec{v}} \equiv \sum_{\alpha=0}^4 \,
     \varphi^{0}_{\alpha \, \Vec{v}} \, \langle \, \Vec{\varphi}^{0}_{\alpha}
     \,,\, \Vec{\psi} \, \rangle  
   \label{eq:035},
   \end{eqnarray}
   and introduce $Q \equiv 1 - P$ as the projection operator to the
    space complement to Ker$L$.
   
   Multiplying Eq.(\ref{eq:020}) by the inverse matrix of
   $f^{\mathrm{eq}}$,  we have
   \begin{eqnarray}
    \frac{\partial}{\partial t}\big(
     f^{\mathrm{eq}\mbox{-}1} \, \tilde{\Vec{f}}^{(1)} \big) 
     = L \, \big( f^{\mathrm{eq} \, \mbox{-}1} \, \tilde{\Vec{f}}^{(1)} \big) 
     + \big( f^{\mathrm{eq} \, \mbox{-}1} \, \Vec{F} \big)\label{eq:036},
   \end{eqnarray}
where $f^{\mathrm{eq}}_{\Vec{v}\Vec{k}} \equiv
   f^{\mathrm{eq}}_{\Vec{v}} \, \delta_{\Vec{v}\Vec{k}}$.
   The  1st-order solution is readily obtained as
   \begin{eqnarray}
    \tilde{\Vec{f}}^{(1)}(t) = e^{(t - t_0)A} \, \Big[ \Vec{f}^{(1)}(t_0) +
     A^{\mbox{-}1} \, \bar{Q} \, \Vec{F} \Big] + (t - t_0)
     \, \bar{P} \, \Vec{F} - A^{\mbox{-}1} \, \bar{Q} \, \Vec{F},
\label{eq:037}
   \end{eqnarray}
   where 
$\bar{P} \equiv f^{\mathrm{eq}} \, P \, f^{\mathrm{eq}\mbox{-}1}$ 
and
 $\bar{Q} \equiv f^{\mathrm{eq}} \, Q \,   f^{\mathrm{eq}\mbox{-}1}$. 

The first order initial value is now determined so that the
would-be fast mode disappear, which in turn 
gives the deformation of the invariant manifold and thereby
 the 1st-order invariant   manifold.
Thus we have for the the 1st-order solution
   \begin{eqnarray}
    \left\{
     \begin{array}{l}
      \displaystyle{\Vec{f}^{(1)}(t_0) = - A^{\mbox{-}1} \, \bar{Q} \, \Vec{F}},\\[0.2cm]
      \displaystyle{\tilde{\Vec{f}}^{(1)}(t) = (t - t_0) \, \bar{P} \,
      \Vec{F} - A^{\mbox{-}1} \, \bar{Q} \, \Vec{F}}.
     \end{array}
    \right.\label{eq:038}
   \end{eqnarray}
   
   Then the 2nd-order equation reads
   \begin{eqnarray}
    \frac{\partial}{\partial t}\tilde{\Vec{f}}^{(2)} = A \,
     \tilde{\Vec{f}}^{(2)} + (t - t_0) \, \Vec{H} + \Vec{I}\label{eq:039},
   \end{eqnarray}
where
   \begin{eqnarray}
    H_{\Vec{v}} \equiv - \Vec{v}\cdot\Vec{\nabla}\Big[ \bar{P} \, \Vec{F}
     \Big]_{\Vec{v}},\,\,\,\,\,\,
     I_{\Vec{v}} \equiv \Vec{v}\cdot\Vec{\nabla}\Big[ A^{\mbox{-}1} \,
     \bar{Q} \, \Vec{F} \Big]_{\Vec{v}}\label{eq:041}.
   \end{eqnarray}
The solution to this equation  is found to be
   \begin{eqnarray}
    \tilde{\Vec{f}}^{(2)}(t) &=& e^{(t - t_0)A} \, \Big[ \Vec{f}^{(2)}(t_0) 
     + A^{\mbox{-}2} \, \bar{Q} \, \Vec{H} + A^{\mbox{-}1} \, \bar{Q} \, \Vec{I} \Big] \nonumber\\
    & & \hspace{-0.5cm}{}+ \frac{1}{2} \, (t - t_0)^2 \, \Big[ \bar{P} \, \Vec{H} \Big] +
     (t - t_0) \, \Big[ \bar{P} \, \Vec{I} - A^{\mbox{-}1} \, \bar{Q} \,
     \Vec{H} \Big] - \Big[ A^{\mbox{-}2} \, \bar{Q} \, \Vec{H} +
     A^{\mbox{-}1} \, \bar{Q} \, \Vec{I} \Big]\label{eq:042}.
   \end{eqnarray}
   Then  the 2nd-order results are summarized as follows:
   \begin{eqnarray}
    \left\{
     \begin{array}{l}
      \displaystyle{\Vec{f}^{(2)}(t_0) = - \Big[ A^{\mbox{-}2} \, \bar{Q} \,
       \Vec{H} + A^{\mbox{-}1} \, \bar{Q} \, \Vec{I} \Big]},\\[0.2cm]
       \displaystyle{\tilde{\Vec{f}}^{(2)}(t) = \frac{1}{2} \, (t -
       t_0)^2 \, \Big[ \bar{P} \, \Vec{H} \Big] + (t - t_0) \, \Big[
      \bar{P} \, \Vec{I} - A^{\mbox{-}1} \, \bar{Q} \, \Vec{H} \Big] -
      \Big[ A^{\mbox{-}2} \, \bar{Q} \, \Vec{H} + A^{\mbox{-}1} \, \bar{Q} \, \Vec{I} \Big]}.
     \end{array}
    \right.\label{eq:043}
   \end{eqnarray}
   
   As a result of the above order-by-order analysis, the invariant
   manifold and the approximate solution up to the 2nd order are found to be
   \begin{eqnarray}
    \Vec{f}(t_0) &=& \Vec{f}^{\mathrm{eq}} - \epsilon
     \, A^{\mbox{-}1} \, \bar{Q} \, \Vec{F} - \epsilon^2 \, \Big[
     A^{\mbox{-}2} \, \bar{Q} \, \Vec{H} + A^{\mbox{-}1} \, \bar{Q} \,
     \Vec{I} \Big]\label{eq:044},\\
    \tilde{\Vec{f}}(t) &=& \Vec{f}^{\mathrm{eq}} +
     \epsilon \, \Bigg( (t - t_0) \, \bar{P} \, \Vec{F} -
     A^{\mbox{-}1} \, \bar{Q} \, \Vec{F} \Bigg) \nonumber\\
    & & \hspace{-0.5cm}{}+ \epsilon^2 \, \Bigg( \frac{1}{2} \, (t - t_0)^2 \, \Big[
     \bar{P} \, \Vec{H} \Big] + (t - t_0) \, \Big[ \bar{P} \, \Vec{I} -
     A^{\mbox{-}1} \, \bar{Q} \, \Vec{H} \Big] - \Big[ A^{\mbox{-}2} \,
     \bar{Q} \, \Vec{H} + A^{\mbox{-}1} \,
     \bar{Q} \, \Vec{I} \Big] \Bigg)\label{eq:045}.
   \end{eqnarray}
   Notice the appearance of secular terms in (\ref{eq:045}).
   
  \subsection{Procedure 3 : envelope equation or RG equation}
   
   Eq.(\ref{eq:045}) shows that the local approximate solution moves away
   from the invariant manifold as $\vert t-t_0\vert$ becomes large
  owing to the secular terms. The appearance of the secular terms invalidates
the perturbation expansion of the solution around $t\simeq t_0$.
   This evolution is described by the microscopic time
described by the kinetic equation Eq.(\ref{eq:008}).
  We can obtain the global solution valid in a global domain
by constructing the envelope of these diverging local solutions parametrized
by $t_0$.
   The envelope equation or the RG equation reads:
   \begin{eqnarray}
    \frac{\partial}{\partial t_0}\tilde{\Vec{f}}(t)\Bigg|_{t_0 = t} = \Vec{0} \,\,\,\,\,\,
     \mathrm{or} \,\,\,\,\,\, \frac{\partial}{\partial
     t_0}\tilde{f}_{\Vec{v}}(\Vec{r} ,\, t \,;\, t_0)\Bigg|_{t_0 = t} = 0\label{eq:046},
   \end{eqnarray}
   which is reduced to
   {\small
   \begin{eqnarray}
    0 = \dot{f}^{\mathrm{eq}}_{\Vec{v}}
     - \epsilon \, \Big[ \bar{P} \, \Vec{F} \Big]_{\Vec{v}} 
     - \epsilon^2 \, \Big[ \bar{P} \, \Vec{I} - A^{\mbox{-}1} \,
     \bar{Q} \, \Vec{H} \Big]_{\Vec{v}} - \frac{\partial}{\partial t} \Bigg\{
     \epsilon \, \Big[ A^{\mbox{-}1} \, \bar{Q} \, \Vec{F} \Big]_{\Vec{v}}
     + \epsilon^2 \, \Big[ A^{\mbox{-}2} \, \bar{Q} \, \Vec{H} +
     A^{\mbox{-}1} \, \bar{Q} \, \Vec{I} \Big]_{\Vec{v}} \Bigg\}.\label{eq:047}
   \end{eqnarray}
   }
   This RG equation is an equation of motion governing the time evolution
   of the five slow variables $n(\Vec{r} ,\,t)$, $T(\Vec{r} ,\,t)$ and
   $\Vec{u}(\Vec{r} ,\,t)$ in $\Vec{f}^\mathrm{eq}$.
   If this equation is solved exactly and these variables are
   substituted in (\ref{eq:044}) at $t_0 = t$, we can construct the macroscopic time
   evolution of the one-particle distribution function $\Vec{f}(t)$.
   In the discussion below we reduce the master equation
   (\ref{eq:047}) to a five-dimensional coupled equation.
   Applying the projection operator $\bar{P}$ from the left of
   (\ref{eq:047}) we obtain the following:
   \begin{eqnarray}
    0 = \Big[ \bar{P} \, \dot{\Vec{f}}^{\mathrm{eq}} \Big]_{\Vec{v}} 
     - \epsilon \, \Big[ \bar{P} \, \Vec{F} \Big]_{\Vec{v}} 
     - \epsilon^2 \, \Big[ \bar{P} \, \Vec{I} \Big]_{\Vec{v}}
     - \Big[ \bar{P} \, \frac{\partial}{\partial t} \Big( 
     \epsilon \, A^{\mbox{-}1} \, \bar{Q} \, \Vec{F} 
     + \epsilon^2 \, \big(
     A^{\mbox{-}2} \, \bar{Q} \, \Vec{H} + A^{\mbox{-}1} \, \bar{Q} \, \Vec{I} \big)
     \Big) \Big]_{\Vec{v}}\label{eq:048}.
   \end{eqnarray}
We notice that the time-derivative does hit to the linear operator $A$ 
which depends on the zero-th order distribution function $\Vec{f}^\mathrm{eq}$,
 which case is in contrast to that treated in the previous section; we remark
that this point was not fully recognized in \cite{hatta}.

Now multipling Eq.(\ref{eq:048}) by the zero modes
   $\varphi^{0}_{\alpha \, \Vec{v}}$ and summing up it in terms of
   $\Vec{v}$, we have
   \begin{eqnarray}
    0 = \sum_{\Vec{v}} \, \varphi^{0}_{\alpha \, \Vec{v}}
     \, \dot{f}^{\mathrm{eq}}_{\Vec{v}} 
     - \epsilon \, \sum_{\Vec{v}} \, \varphi^{0}_{\alpha \, \Vec{v}} \, F_{\Vec{v}} 
     - \epsilon^2 \, \sum_{\Vec{v}} \, \varphi^{0}_{\alpha \, \Vec{v}}
     \, I_{\Vec{v}}
     \,\,\, \mathrm{for} \,\,\, \alpha = 0 ,\, 1 ,\, 2 ,\, 3 ,\, 4\label{eq:049}.
   \end{eqnarray}
Here we have used the following relations
obtained from the definitions (\ref{eq:024}) and (\ref{eq:035}),
   \begin{eqnarray}
    \sum_{\Vec{v}} \, \varphi^{0}_{\alpha \, \Vec{v}} \, \Big[ \bar{P} \,
     \psi \Big]_{\Vec{v}} = \sum_{\Vec{v}} \, \varphi^{0}_{\alpha \, \Vec{v}} \,
     \psi_{\Vec{v}},\,\,\,\,\,\,
     \sum_{\Vec{v}} \, \varphi^{0}_{\alpha \, \Vec{v}} \, \Big[ \bar{Q} \,
     \psi \Big]_{\Vec{v}} =  0\label{eq:051},
   \end{eqnarray}
   and the equality
   \begin{eqnarray}
    \sum_{\Vec{v}} \, \varphi^{0}_{\alpha \, \Vec{v}} \,
     \frac{\partial}{\partial t} 
     \Bigg\{ 
     \epsilon \, \Big[ A^{\mbox{-}1} \, \bar{Q} \, \Vec{F} \Big]_{\Vec{v}}
     + \epsilon^2 \, \Big[ A^{\mbox{-}2} \, \bar{Q} \, \Vec{H} +
     A^{\mbox{-}1} \, \bar{Q} \, \Vec{I} \Big]_{\Vec{v}} \Bigg\} = 0
\label{eq:052},
   \end{eqnarray}
which follows from the fact that
   $\dot{\Vec{\varphi}}^{0}_{\alpha} \in \mathrm{Ker} \, L$.
   Notice that (\ref{eq:049}) has the same information as the
   RG equation (\ref{eq:047}).
   
\subsection{Explicit reduction to Navier-Stokes equation; transport coefficients}   
   
   We now show that 
   Eq.(\ref{eq:049}) is nothing but the Navier-Stokes equation\cite{resibois}.
   Performing the summation in terms of $\Vec{v}$, we find that the
   first and second terms of Eq.(\ref{eq:049}) are evaluated to be
   {\small
   \begin{eqnarray}
    \sum_{\Vec{v}} \, \varphi^{0}_{0 \, \Vec{v}} \, \dot{f}^{\mathrm{eq}}_{\Vec{v}} 
     &=& \Big( \frac{1}{\sqrt{n}} \, \frac{1}{m} \big) \,m \, \dot{n}\label{eq:053},\\
    \sum_{\Vec{v}} \, \varphi^{0}_{i \, \Vec{v}} \, \dot{f}^{\mathrm{eq}}_{\Vec{v}} 
     &=& \Big( \frac{1}{\sqrt{n}} \, \sqrt{\frac{m}{T}} \, \frac{1}{m} \Big) \, m \, n \,
     \dot{u}^i \,\,\, \mathrm{for} \,\,\, i = 1 ,\, 2 ,\, 3\label{eq:054},\\
    \sum_{\Vec{v}} \, \varphi^{0}_{4 \, \Vec{v}} \, \dot{f}^{\mathrm{eq}}_{\Vec{v}} 
     &=& \Big( \frac{1}{\sqrt{n}} \, \sqrt{\frac{2}{3}} \, \frac{1}{T} \Big)
     \, n \, \frac{3}{2} \, \dot{T}\label{eq:057},
   \end{eqnarray}
   }
   and
   {\small
   \begin{eqnarray}
    \sum_{\Vec{v}} \, \varphi^{0}_{0 \, \Vec{v}} \, F_{\Vec{v}} 
     &=& - \Big( \frac{1}{\sqrt{n}} \, \frac{1}{m} \big) \, 
     m \, \Vec{\nabla}\cdot(n \, \Vec{u})\label{eq:058},\\
    \sum_{\Vec{v}} \, \varphi^{0}_{i \, \Vec{v}} \, F_{\Vec{v}} 
     &=& - \Big( \frac{1}{\sqrt{n}} \, \sqrt{\frac{m}{T}} \, \frac{1}{m} \Big) \, 
     \Big( m \, n \, \Vec{u}\cdot\Vec{\nabla}u^i + \nabla^i(n \, T)
     \Big) \,\,\, \mathrm{for} \,\,\, i = 1 ,\, 2 ,\, 3\label{eq:059},\\
    \sum_{\Vec{v}} \, \varphi^{0}_{4 \, \Vec{v}} \, F_{\Vec{v}} 
     &=& - \Big( \frac{1}{\sqrt{n}} \, \sqrt{\frac{2}{3}} \, \frac{1}{T} \Big)
     \, \Big( n \, \Vec{u}\cdot\Vec{\nabla}\big( \frac{3}{2} \, T \big) + n \, T \,
     \Vec{\nabla}\cdot\Vec{u} \Big)\label{eq:062}.
   \end{eqnarray}
   }
   In a similar way, the third terms of Eq.(\ref{eq:049}) are evaluated to be
   {\small
   \begin{eqnarray}
    \hspace{-0.5cm}\sum_{\Vec{v}} \, \varphi^{0}_{0 \, \Vec{v}} \, I_{\Vec{v}} 
     &=& 0\label{eq:063},\\
    \hspace{-0.5cm}\sum_{\Vec{v}} \, \varphi^{0}_{i \, \Vec{v}} \, I_{\Vec{v}} 
     &=& \Big( \frac{1}{\sqrt{n}} \, \sqrt{\frac{m}{T}} \, \frac{1}{m} \Big) \, 
     \nabla^j \langle \, \Vec{T}^{ij} \,,\, L^{\mbox{-}1} \, Q \,
     f^{\mathrm{eq}\mbox{-}1} \, \Vec{F} \, \rangle \,\,\, \mathrm{for}
     \,\,\, i = 1 ,\, 2 ,\, 3\label{eq:064},\\
    \hspace{-0.5cm}\sum_{\Vec{v}} \, \varphi^{0}_{4 \, \Vec{v}} \, I_{\Vec{v}} 
     &=& \Big( \frac{1}{\sqrt{n}} \, \sqrt{\frac{2}{3}} \, \frac{1}{T} \Big) \, 
     \Big( \nabla^i \langle \, \Vec{J}^{i} \,,\, L^{\mbox{-}1} \, Q \,
     f^{\mathrm{eq}\mbox{-}1} \, \Vec{F} \, \rangle 
     + \langle \, \Vec{T}^{ij} \,,\, L^{\mbox{-}1} \, Q \,
     f^{\mathrm{eq}\mbox{-}1} \, \Vec{F} \, \rangle \,
     \frac{1}{2} \, ( \nabla^i u^j + \nabla^j u^i ) \Big)\label{eq:067},
   \end{eqnarray}
   }
   where $\Vec{T}^{ij}$ and $\Vec{J}^{i}$ are defined as
   \begin{eqnarray}
    T^{ij}_{\Vec{v}} \equiv m \, \delta v^i \, \delta v^j,\,\,\,\,\,\,
     J^{i}_{\Vec{v}} \equiv \Big( \frac{1}{2} \, m \,
     |\Vec{\delta v}|^2 - \frac{5}{2} \, T \Big) \, \delta v^i\label{eq:069}.
   \end{eqnarray}
   Here  we have replaced
   $\varphi^{0}_{4 \, \Vec{v}} \, \delta v^i$ by $J^{i}_{\Vec{v}}$
   because $\delta v^i$ belongs to $\mathrm{Ker} \, L$ and its inner
   product with $\Big[ L^{\mbox{-}1} \, Q \, f^{\mathrm{eq} \, \mbox{-}1} \, \Vec{F}
   \Big]_{\Vec{v}}$ is null.

   To proceed further, the explicit  representation of
   $\Big[ L^{\mbox{-}1} \, Q \, f^{\mathrm{eq} \, \mbox{-}1} \, \Vec{F}
   \Big]_{\Vec{v}}$ is necessary. By using the projection operator
   (\ref{eq:035}), we have
   \begin{eqnarray}
    \Big[ L^{\mbox{-}1} \, Q \, f^{\mathrm{eq}\mbox{-}1} \, \Vec{F}
     \Big]_{\Vec{v}} = - \frac{1}{T} \, \sum_{\Vec{k}} \,
     L^{\mbox{-}1}_{\Vec{v}\Vec{k}} \, \Bigg[ T^{ij}_{\Vec{k}} \,
     \frac{1}{2} \, \Big( \nabla^i u^j + \nabla^j u^i - \frac{2}{3} \, \delta^{ij} \,
     \Vec{\nabla}\cdot\Vec{u} \Big) + J^{i}_{\Vec{k}} \, \nabla^{i} \ln T \Bigg]\label{eq:070}.
   \end{eqnarray}
   Following the above equation and the space rotational
   symmetry\cite{chapman, kawasaki}, we arrive at
   \begin{eqnarray}
    \langle \, \Vec{T}^{ij} \,,\, L^{\mbox{-}1} \, Q \,
     f^{\mathrm{eq}\mbox{-}1} \, \Vec{F} \, \rangle &=&  2 \, \eta \,
     \frac{1}{2} \, \Big( \nabla^i u^j + \nabla^j u^i - \frac{2}{3} \,
     \delta^{ij} \, \Vec{\nabla}\cdot\Vec{u} \Big)\label{eq:071},\\
    \langle \, \Vec{J}^{i} \,,\, L^{\mbox{-}1} \, Q \,
     f^{\mathrm{eq}\mbox{-}1} \, \Vec{F} \, \rangle &=& \lambda \, \nabla^iT\label{eq:072},
   \end{eqnarray}
   where $\eta$ and $\lambda$ are the so-called transport coefficients
   defined by
   {\small
   \begin{eqnarray}
    & &\hspace{-0.5cm}\eta \equiv - \frac{1}{T} \, \sum_{\Vec{v}\Vec{k}} \,
     f^{\mathrm{eq}}_{\Vec{v}} \, T^{12}_{\Vec{v}} \,
     L^{\mbox{-}1}_{\Vec{v}\Vec{k}} \, T^{12}_{\Vec{k}}
     = \frac{1}{T} \, \int_0^\infty\!\!d\tau \, \langle \, \Vec{T}^{12}(0)
     \,,\, \Vec{T}^{12}(\tau) \, \rangle,\,\,\,
     \Big[ \Vec{T}^{ij}(\tau) \Big]_{\Vec{v}} \equiv \sum_{\Vec{k}} \, \Big[
     e^{L\tau} \Big]_{\Vec{v}\Vec{k}} \, T^{ij}_{\Vec{k}}\label{eq:073},\\
    & &\hspace{-0.5cm}\lambda \equiv - \frac{1}{T^2} \, \sum_{\Vec{v}\Vec{k}} \,
     f^{\mathrm{eq}}_{\Vec{v}} \, J^{1}_{\Vec{v}} \,
     L^{\mbox{-}1}_{\Vec{v}\Vec{k}} \, J^{1}_{\Vec{k}} 
     = \frac{1}{T^2} \, \int_0^\infty\!\!d\tau \, \langle \, \Vec{J}^{1}(0)
     \,,\, \Vec{J}^{1}(\tau) \, \rangle,\,\,\,
     \Big[ \Vec{J}^{i}(\tau) \Big]_{\Vec{v}} \equiv \sum_{\Vec{k}} \, \Big[
     e^{L\tau} \Big]_{\Vec{v}\Vec{k}} \, J^{i}_{\Vec{k}}\label{eq:074}.
   \end{eqnarray}
   }
   We note that the transport coefficients, shear
   viscosity $\eta$ and heat conductivity $\lambda$, are obtained as the
   correlation function of the microscopic currents (\ref{eq:069}) like
   Kubo formula\cite{kubo}.
   
   Putting back $\epsilon = 1$, 
   the reduced RG equations (\ref{eq:049}) are found to be
   \begin{eqnarray}
    \dot{n} + \Vec{\nabla}\cdot(n \, \Vec{u}) &=& 0\label{eq:075},\\
    m \, n \, \dot{u}^i + m \, n \, \Vec{u}\cdot\Vec{\nabla}u^i &=& -
     \nabla^j \, P^{ji} \,\,\, \mathrm{for} \,\,\, i = 1 ,\, 2 ,\, 3\label{eq:076},\\
    n \, \dot{e} + n \, \Vec{u}\cdot\Vec{\nabla}\epsilon &=& -
     \Vec{\nabla}\cdot\Vec{J} - P^{ij} \, D^{ij}\label{eq:077},
   \end{eqnarray}
   where
   \begin{eqnarray}
    & & e \equiv \frac{3}{2} \, T,\,\,\,\,\,\,
     p \equiv n \, T,\,\,\,\,\,\,
     D^{ij} \equiv \frac{1}{2}(\nabla^i u^j + \nabla^j u^i)\label{eq:082},\\
    & & P^{ij} \equiv \delta^{ij} \, p - 2 \, \eta \, D^{ij} - \big(
     - \frac{2}{3} \, \eta \big) \, \delta^{ij} \,
     \Vec{\nabla}\cdot\Vec{u},\,\,\,\,\,\,
     \Vec{J} \equiv - \lambda \, \Vec{\nabla}T\label{eq:079}.
   \end{eqnarray}
   These equations (\ref{eq:075})-(\ref{eq:077}) is identified with
   the fluid dynamic equations with dissipation, i.e., the
   Navier-Stokes equation\cite{resibois}. 

In summary, we have shown that the 
   Navier-Stokes equations can be neatly reproduced as the fluid dynamical
   limit of the Boltzmann equation in the RG method.
   
\section{Summary and concluding remarks}

We have shown that the RG gives a powerful and systematic method for the reduction
of the dynamics and also provides a transparent way for the construction
of the attractive slow manifold.
We have indicated the relation of the underlying
mathematics of the RG method with the classical theory of envelopes in
mathematical analysis.
 Although the uses of envelopes for physics
problem was first noted by M. Suzuki in the theory of 
CAM(Coherent Anomaly Method)\cite{cam} for the critical phenomena 
in statistical physics, the usefulness of the notion of 
envelopes was not well recognized.

We have obtained the Navier-Stokes equation from
the Boltzmann equation by applying the RG method:
We have worked out for constructing the projection operators
and thus explicitly given the forms of the transport coefficients in terms
of the one-particle distribution function.

The RG method presented here has a wide range of applicabilities even
being confined to the transport equations\cite{hatta}:
It can be applied to obtain the Boltzmann equation from the Liouville equation.
  The Focker-Planck equation is equally 
obtained from the Langevin equation by this method.
Furthermore, the further reduction of the Focker-Planck equation can be also done by the
RG method.

As for the reduction of the hydrodynamic equation 
from  the Boltzmann equation, it would be interesting
to apply the present method to the relativistic case.
We also mention that the present method is also applicable to extract the critical
slow dynamics around the critical point of 
phase transitions\cite{kunihiro3,efk}. It would be also
interesting to apply the method for extracting the slow dynamics,
 say, around the QCD critical end point\cite{cep}.

\section*{Acknowledgments}

T. K. is supported by Grant-in-Aide for Scientific Research by
   Monbu-Kagaku-sho (No.\ 17540250).
This work is supported in part by a Grant-in-Aid for the 21st Century 
   COE ``Center for Diversity and Universality in Physics''.

\input{ref.tex}

\end{document}

%% file: ref.tex
\newcommand{\NGold}{N. \ Goldenfeld}
\newcommand{\YO}{Y.\ Oono}